\documentclass[aps,
prd,twocolumn,amsmath,amssymb]{revtex4}
\usepackage{color}
\usepackage{graphics}
\usepackage{epsfig}
\usepackage{amsmath}
\usepackage{amssymb}
\usepackage{amsthm}
\usepackage[mathscr]{eucal}

\newcommand{\bea}{\begin{eqnarray}}
\newcommand{\eea}{\end{eqnarray}}
\newcommand{\beas}{\begin{eqnarray*}}
\newcommand{\eeas}{\end{eqnarray*}}

\begin{document}

\title{Dual symmetries of dense isotopically and chirally asymmetric QCD}
\author{K. G. Klimenko and R. N. Zhokhov }
\affiliation{NRC "Kurchatov Institute"--IHEP, Protvino, 142281, Russian Federation}

\begin{abstract}
In the present paper, the dual symmetries of dense quark matter  phase diagram found in 
some massless three- and two-color NJL  models in the mean field approximation have
been shown to exist at a more fundamental level as dual transformations of fields and 
chemical potentials leaving the Lagrangian invariant. As a result, the corresponding dual 
symmetries of the full phase diagram can be shown without any approximation. 
And it has been shown not only in the NJL models, but also in framework of two- and three-color massless 
QCD itself. This is quite interesting, since one might say that it is not very common to show something completely non-perturbatively in QCD.
\end{abstract}
	
\maketitle

{\bf Introduction.---}At present, there is a reasonable confidence that 
dense quark matter can be formed in heavy ion collisions or in the cores of neutron stars. 
The quantum chromodynamics (QCD), effective Nambu-Jona-Lasinio (NJL) models, etc
\cite{buballa2,asakawa,alford}, are usually used to describe the physical processes occurring
in it. It is evident that the properties of cold quark matter are largely determined by the 
values of its baryon $n_B$ and isospin $n_I$ densities. In addition, if hot and/or dense quark matter is under
the influence of a strong magnetic field or rotation, its phase structure can also depend on the chiral 
$n_5$ and chiral-isospin $n_{I5}$ densities. These densities can arise in the quark medium in real physical scenarios (e. g. due to chiral separation effect \cite{Vilenkin,Shovkovy} and in strong electromagnetic fields \cite{Ruggieri:2016lrn} (see discussion in \cite{kkz20, khun2, khun} and references therein)) and lead to the chiral magnetic \cite{Fukushima}.
Depending on the relationship between these densities (or, equivalently, between the 
corresponding thermodynamically conjugate chemical potential values $\mu_B$, $\mu_I$, $\mu_5$ and $\mu_{I5}$) in dense quark medium formed only by $u$ and $d$ quarks the color 
superconducting \cite{alford,u2}, charged pion condensation  
\cite{son,ak,he,ekkz2} and the phases with chiral and chiral-isospin asymmetries 
\cite{andrianov,khun,chao,ramos}, etc. can be observed.
At the same time, it becomes obvious that the greater 
the number of chemical potentials, and therefore exotic phases, are involved in  
considering the properties of the phase structure of dense quark matter, the more and more complicated 
this task becomes even within the framework of the simplest analytical approaches, not 
to mention the fundamental lattice QCD approach to the problem. However, one circumstance 
was recently discovered that suggests that the task of studying the phase structure of 
real dense quark matter is not so hopeless. 

Indeed, within the framework of the simplest 
NJL model, it was discovered that in the mean field approximation 
(or in the leading order of the $1/N_c$ expansion, where $N_c$ is the number of colors) 
the chiral symmetry breaking (CSB) and 
charged pion condensation (PC) phases are dually conjugate to each other \cite{kkz18,kkz18-2}.
It means that there is symmetry of the phase diagram with respect to the rearrangement of 
the order parameters, namely of chiral and charged pion condensates, as well as 
the simultaneous transformation of chemical potentials $\mu_I\leftrightarrow\mu_{I5}$. 
Let us note that duality was found in the case when current quark mass is zero, i.e. in 
the chiral limit. However, at the physical point, i.e. at physical values of (nonzero) 
current quark masses, duality is approximate, but it holds with very good approximation and 
quite robust under influence of temperature, etc \cite{khun2}.  Moreover, as it turned out, the duality between CSB and charged PC is not just interesting and beautiful mathematical feature of QCD phase diagram, which is 
undoubtedly true, but a rather useful tool of investigating phase diagram or at least 
facilitating its research. For example, using only the duality mapping and known 
$(\mu_B,\mu_I)$-phase structure of quark matter, its $(\mu_B,\mu_{I5})$-phase portrait 
with new inhomogeneous phases has been obtained without any calculations \cite{kkzparticles}, etc.

The duality between CSB and charged PC  phases has been shown for the first time in the framework of the (1+1)-dimensional massless
two-flavor NJL model in the leading order of $1/N_{c}$  approximation  \cite{k} (see
also in \cite{kkz}). The possibility for duality correspondence between other physical 
phenomena, e.g.,  CSB and superconductivity (diquark condensation), has been also investigated earlier in the 
framework of low dimensional quantum field theories with four-fermion interactions in 
\cite{thies,ebert,cao}. 

Finally, note that in dense isotopically and chirally asymmetric baryonic matter, which is formed by two-color quarks and described by the massless two-flavor and two-color NJL model, two more dual symmetries of its mean-field thermodynamic potential (TDP) arise \cite{kkz20}.
As a result, the dualities between CSB and baryon superfluid (BSF), or diquark condensation, phases as well as between charged PC and BSF phenomena appear, in addition to the above-mentioned duality between CSB and charged PC. In this NJL model, 
which is a low-energy effective model of the two-color QCD with two flavors, the investigation of dense quark matter was also 
performed only in the mean-field approximation.

Moreover, in a more involved three-color and two-flavor NJL model with additional 
diquark interaction channel \cite{u2} the same dual symmetry between CSB and charged PC phases 
have been observed in the chiral limit and also using the mean-field approximation as in the 
simplest NJL model \cite{kkz18}. Hence, we see that two different three-color NJL models 
that effectively describe (in the general case, different) low-energy QCD regions have TDPs 
that are invariant in the chiral limit and in the mean-field 
approximation under the same dual transformation relating the phenomena CSB and charged PC.

Now, two questions naturally arise. (i) Whether the dual relations between various physical phenomena are inherent in 
the above-mentioned dense NJL models as a whole, and not only in their mean-field 
approximations. In addition, (ii) one can pose the question even more 
broadly and try to clarify the situation with the dual symmetries within dense QCD 
itself, both with two and three colors of quarks. The present paper is devoted to the 
consideration of this questions. 

We show that massless QCD and QCD-like effective NJL Lagrangians 
are invariant with respect to some duality transformations. Due to this fact, the confidence
appears that the corresponding full TDPs are dually symmetric, i.e. the dualities between 
different phenomena, which were previously observed in the mean-field or large-$N_c$ approximations, are inherent in the QCD or 
the corresponding NJL models itself, on the basis of which dense quark matter is studied.

{\bf The case of two-color models.---}Our starting point is the Lagrangian for 
the quark sector of two-color massless QCD extended by four chemical potential terms 
\begin{equation}
L_{QC_2D}=i\bar\psi \gamma^{\mu}D_{\mu} \psi+ \bar\psi {\cal M} \psi,
\label{1}
\end{equation} 
where $
{\cal M}=\frac{\mu_B}{2}\gamma^{0}+\frac{\mu_{I}}{2}
\gamma^{0}\tau_{3}+\frac{\mu_{I5}}{2}\gamma^{0}\gamma^{5}\tau_{3}+\mu_5\gamma^{0}\gamma^5$. In (\ref{1}) and below, quark field $\psi(x)$ is a flavor doublet, i.e. $\psi^T=
(u^T,d^T)$, and each of $u,d$ is a color doublet and four-component Dirac spinor. 
$D_{\mu}=\partial_\mu-ig\sigma_aA^a_\mu(x)$ (here $\sigma_a$ ($a=1,2,3$) are three 2$\times$2
color Pauli matrices). As it was shown in \cite{Kogut,SonSplittorff}, at ${\cal M}=0$ the 
Lagrangian (\ref{1}) is invariant under $SU(4)$ symmetry group, since in this case 
it can be presented in 
the form $L_{QC_2D}=i\bar\Psi \gamma^{\mu}D_{\mu} \Psi$, where $\Psi(x)$ is an auxiliary
spinor field, $\Psi^{T} = \left( u^T_L,d^T_L, \sigma_{2}(u_{R}^{C})^{T}, \,
\sigma_{2} (d_{R}^{C})^{T}	 \right),\;
\bar\Psi = \left( \bar u_{L}, 	\bar d_{L}, \bar u_{R}^{C}\sigma_{2},
\bar d_{R}^{C}\sigma_{2} \right)$, and it belongs to fundamental representation of SU(4) group.

Let us now consider two-color NJL model. Its kinetic term explicitly invariant under
$SU(4)$ group could be easily written as $i\bar\Psi \gamma^{\mu}\partial_{\mu} \Psi$. 
The invariant with respect to $SU(4)$ interaction terms of the 2-color NJL Lagrangian 
can be constructed from the two $SU(4)$-invariant structures,
$|\bar \Psi^{C} \vec\Sigma \Psi|^{2}$ and $(\bar \Psi^{C} \vec\Sigma \Psi)^{2} + 
{\rm H.c.}$ (for detail, see in \cite{Andersen}), where 
$\vec\Sigma$ are explicitly:
{\small 
$\Sigma_{1} = \left( {\begin{array}{cccc}
0 & -1 \\
1  & 0
\end{array} }\right),\; \Sigma_{2} = \left( {\begin{array}{cccc}
\tau_{2} & 0 \\
0  & \tau_{2}
\end{array} } \right),\;\Sigma_{3} = \left( {\begin{array}{cccc}
0 & 	i\tau_{1} \\
 -i\tau_{1}  & 0
\end{array} } \right),\;\Sigma_{4} = \left( {\begin{array}{cccc}
i\tau_{2} & 0 \\
0  & -i\tau_{2}
\end{array} } \right),\;
\Sigma_{5} = \left( {\begin{array}{cccc}
0 & 	i\tau_{2} \\
-i\tau_{2}  & 0
\end{array} } \right),\;\Sigma_{6} = \left( {\begin{array}{cccc}
0 & 	i\tau_{3} \\
-i\tau_{3}  & 0
\end{array} } \right)
$.} Note that 
$\bar \Psi^{C} \vec\Sigma \Psi$ transforms as a fundamental representation of SO(6) group. 
Moreover, there is an isomorphism $SU(4)/Z_2=SO(6)$, i.e. up to $Z_2$ these groups, $SU(4)$ and $SO(6)$, are isomorphic. 
Next, we will consider the simplest version of the $SU(4)$ symmetric 
QC$_2$D-like effective four-fermion model, which in terms of $\psi(x)$ fields has the following Lagrangian
\vspace{-0.3cm}
\begin{eqnarray}
&& \widetilde L_{NJL_2}= i\bar{\psi} \gamma^{\mu}\partial_{\mu} \psi+ G\big[(\bar\psi \psi)^{2}    \nonumber\\
&& +(i\bar\psi \vec \tau \gamma^{5} \psi)^{2}+(i\bar\psi \sigma_{2}\tau_{2}\gamma^{5} \psi^{C})(i\bar\psi^{C} \sigma_{2}\tau_{2}\gamma^{5} \psi)\big].\label{10}
\end{eqnarray}
In order to study dense quark matter in the framework of this 2-color NJL model, one should add to the Lagrangian (\ref{10}) the chemical potential term $\bar\psi{\cal M}\psi$ (see in (\ref{1})), i.e. consider $L_{NJL_2}\equiv\widetilde L_{NJL_2}+\bar\psi{\cal M}\psi$.
This term breaks $SU(4)$ symmetry, but it is quite insightful 
to rewrite it with the use of auxiliary spinor fields $\Psi$,\vspace{-1.3cm}
\begin{widetext}
\begin{equation}
\bar\psi{\cal M}\psi=\frac{\mu_B}2\Psi^{\dagger} \left( {\begin{array}{cccc}
1 & 0 \\
0  & -1
\end{array} } \right) \Psi+\frac{\mu_{I}}{2}\Psi^{\dagger} \left( {\begin{array}{cccc}
\tau_{3} & 0 \\
0  & -\tau_{3}
\end{array} } \right) \Psi+\frac{\mu_{I5}}{2}\Psi^{\dagger} \left( {\begin{array}{cccc}
\tau_{3} & 0 \\
0  & \tau_{3}
\end{array} } \right) \Psi+\mu_{5}\Psi^{\dagger} \left( {\begin{array}{cccc}
1 & 0 \\
0  & 1
\end{array} } \right) \Psi.
 \label{13}
\end{equation}
\end{widetext}
$\bullet$ Now having in $L_{QC_2D}$ and $L_{NJL_2}$ terms with chemical 
potentials written in the form (\ref{13}), one can notice that if we 
change the order of the second and third components in $\Psi$ field, i.e. make the 
transformation $d_{L}\leftrightarrow\sigma_{2}u_{R}^{C}$, then the fermion structure 
of terms with chemical potential $\mu_{I5}$ and $\mu_B$, i.e. first and the third one 
in (\ref{13}), transforms one into another. And if we in addition transform the chemical 
potential $\mu_B\leftrightarrow\mu_{I5}$ then these terms stays intact.
Besides, one can show that other terms with chemical potentials $\mu_I$ and $\mu_5$ are invariant with respect to this transformation. 
Kinetic and interaction parts of Lagrangians are also invariant. One could write this (dual) transformation in the following matrix form 
\begin{equation}
{\cal D_{\rm III}}: \left( {\begin{array}{cc}
d_{L} \\
\sigma_{2}u_{R}^{C} \\
\end{array} } \right)\to i\tau_{2}\left( {\begin{array}{cc}
d_{L} \\
\sigma_{2}u_{R}^{C} \\
\end{array} } \right);~
\mu_B\leftrightarrow\mu_{I5}. \label{03}
\end{equation}
$\bullet$ Let us see that if we change the order of first and third components in $\Psi$ field
then the structure of terms with chemical potentials $\mu_{I}$ and $\mu_B$ (first and 
second term in (\ref{13})) transforms one into the other. And if, in addition, one make the 
transformation $\mu_B\leftrightarrow\mu_I$, then both these terms and the Lagrangians 
$L_{QC_2D}$ and $L_{NJL_2}$ as a whole will remain unchanged.
This dual transformation has the form
\begin{equation}
{\cal D_{\rm II}}:
\left( {\begin{array}{cc}
d_{L} \\
\sigma_{2}d_{R}^{C} \\
\end{array} } \right)\to
i\tau_{1}\left( {\begin{array}{cc}
d_{L} \\
\sigma_{2}d_{R}^{C} \\
\end{array} } \right);~\mu_B\leftrightarrow\mu_I. \label{003}
\end{equation}
$\bullet$ Now let us further observe that if we change last two components of the $\Psi$ field 
and make at the same time a rearrangement $\mu_I\leftrightarrow\mu_{I5}$, then two-color 
Lagrangians (\ref{1}) and (\ref{10}) will also remain unchanged. In the matrix form this dual transformation looks like
\begin{equation}
{\cal D_{\rm I}}:
\left( {\begin{array}{cc}
u_{R} \\
d_{R}
\end{array} } \right)\to
i\tau_1\left( {\begin{array}{cc}
u_{R} \\
d_{R} \\
\end{array} } \right);~\mu_I\leftrightarrow\mu_{I5}.\label{5}
\end{equation}
We emphasize that during dual transformations, not only the field variables change, 
but also the free parameters of the model (in our case, these are chemical potentials). 
Whereas with conventional symmetry transformations of the model, only the fields vary.

Thus, we have shown that at ${\cal M}\ne 0$ the Lagrangians of the two-color QCD and NJL 
models are invariant under the dual transformations (\ref{03})-(\ref{5}). Since in this case 
discrete field transformations are from $SU(4)$ group (which is not anomalous), 
the dualities are not to be broken by the anomaly.

As a consequence, the full TDP $\Omega$ of any of the QCD and QCD-like models also has this property, which 
is not difficult to give a more physical interpretation. Indeed, the TDP depends both on 
chemical potentials $\mu_B$,... and on a number of order parameters. In our case these are 
the following ground state expectation values: \begin{eqnarray}
\sigma\equiv\langle
\bar\psi \psi\rangle, \vec\pi\equiv \langle\bar\psi i\gamma^5\vec\tau\psi
\rangle,
\delta\equiv\langle\psi^T
Ci\gamma^5\sigma_2\tau_2\psi \rangle.
\label{07}
\end{eqnarray}
However, in the chiral limit the number of order 
parameters is reduced, and the full TDP is indeed a function vs $M,\Pi$ and $\Delta$, where  
$M=\sqrt{\sigma^2+\pi^2_0}$, $\Pi=\sqrt{\pi_1^2+\pi_2^2}$, $\Delta=\sqrt{\delta^*\delta}$ 
(for details, see in \cite{kkz20}), i.e. $\Omega=\Omega(\mu_B,...;M,\Pi,\Delta)$. Note that the form of the global minimum point (GMP) of this TDP vs 
$M,\Pi,\Delta$ defines the phase structure of the model. So if the GMP looks like $(M_0,0,0)$, then
CSB phase is realized in the model, if it has the form $(0,\Pi_0,0)$ -- we have charged PC 
phase, and the GMP of the form $(0,0,\Delta_0)$ corresponds to the BSF phase \cite{kkz20}.

Generally speaking, construction of a phase portrait in the above-mentioned 2-color models 
is a rather difficult task. But it is significantly simplified if we take into account 
the dual symmetries (\ref{03})-(\ref{5}) of the Lagrangians (\ref{1}) and (\ref{10}). As a consequence, the complete TDP of each of the models must be invariant under the following transformations: 
(i) ${\cal D_{\rm III}}:~M\leftrightarrow\Delta,~\mu_B\leftrightarrow\mu_{I5}$, (ii) ${\cal D_{\rm II}}:~\Delta\leftrightarrow\Pi,~\mu_B\leftrightarrow\mu_{I}$, (iii) ${\cal D_{\rm I}}:~M\leftrightarrow\Pi,~\mu_I\leftrightarrow\mu_{I5}$.
It means that if at some chemical potential point $(\mu_B=A,\mu_I=B,\mu_{I5}=C,\mu_5)$ we have, e.g., the CSB phase, then at the point $(\mu_B=C,\mu_I=B,\mu_{I5}=A,\mu_5)$ the BSF phase is realized, whereas at the point $(\mu_B=A,\mu_I=C,\mu_{I5}=B,\mu_5)$ the charged PC is placed, etc. 

{\bf Three-color models.---}Let us consider duality invariance of the massless 2-flavor QCD and corresponding 
NJL effective models when $N_c=3$. In this case the QCD Lagrangian $L_{QCD}$ has the form 
(\ref{1}), where $D_{\nu}=\partial_{\nu}+ig\lambda_{a}A^{a}_{\nu}$ (summation over $a=1,..,8$ is implied),
and minor replacement $\mu_B/2\to\mu_B/3$ should be done. Dual properties of the simplest
NJL model,
\begin{eqnarray}
&&L_{NJL}=
\bar{\psi}\gamma^{\mu}\partial_{\mu}\psi+\bar{\psi}{\cal M} \psi
+G\big\{(\bar\psi \psi)^{2}+(i\bar\psi \vec \tau \gamma^{5} \psi)^{2}\big\}\nonumber\\
&&+
H\sum_{a=2,5,7}(i\bar\psi \lambda_{a}\tau_{2}\gamma^{5} \psi^{C})(i\bar\psi^{C} 
\lambda_{a}\tau_{2}\gamma^{5} \psi),
\label{7}
\end{eqnarray}
are also established here.
Let us note that dualities ${\cal D_{\rm II}}$ and ${\cal D_{\rm III}}$ of the two-color QCD 
are defined with 
the use of the two-color structures (explicitly consist of $\sigma_2$). But if we take a 
look at the ${\cal D_{\rm I}}$ duality (\ref{5}), it does not contain two-color structures 
in its definition, and if one tries to perform this transformation in three-color QCD and NJL 
models, then it could be evident that $L_{QCD}$ and $L_{NJL}$ (\ref{7}) are invariant 
under this transformation. Indeed, some of chemical potential terms are invariant under 
${\cal D_{\rm I}}$, $\bar{\psi}\gamma^{0}\psi\leftrightarrow\bar{\psi}\gamma^{0}
\psi$ and $\bar{\psi}\gamma^{0}\gamma^{5}\psi\leftrightarrow\bar{\psi}\gamma^{0}
\gamma^{5}\psi$. But  the remaining two transform into each other,
$\bar{\psi}\gamma^{0}\gamma^{5}\tau_3\psi\leftrightarrow\bar{\psi}\gamma^{0}
\tau_3\psi$, so one needs $\mu_I\leftrightarrow\mu_{I5}$. Moreover, since the  
bilinear fermion structures of both Lagrangians are transformed by ${\cal D_{\rm I}}$ 
in the following way: $\bar{\psi}\psi\leftrightarrow
i \bar{\psi}\gamma^{5}\tau_{1}\psi$, $i \bar{\psi}\gamma^{5}\tau_{2}\psi\leftrightarrow 
i\bar{\psi}\gamma^{5}\tau_{3}\psi$ and $i\bar\psi^{C} \lambda_{a}\tau_{2}\gamma^{5} 
\psi\leftrightarrow
i\bar\psi^{C} \lambda_{a}\tau_{2}\gamma^{5} \psi$, one can conclude that $L_{QCD}$ and 
$L_{NJL}$ (\ref{7}) are invariant under the duality transformation ${\cal D_{\rm I}}$ (\ref{5}).
The field transformation in (\ref{5}) lies in $SU(2)_{L}\times SU(2)_{R}$ and 
the dual ${\cal D_{\rm I}}$ symmetry can be shown indeed not just for the simplest NJL model 
(\ref{7}), but for any effective QCD model invariant under $SU(2)_{L}\times SU(2)_{R}$. 
Since this group is not anomalous, the duality ${\cal D_{\rm I}}$ is not to be broken by 
anomaly in 3-color QCD and corresponding QCD-like models. 

Notice that the full TDP $\Omega$ in both 3-color models is indeed a function 
of order parameters $\sigma$, $\vec\pi$ (see in (\ref{07})) and $\Delta_a=\langle 
i\bar\psi^{C} \lambda_{a}\tau_{2}\gamma^{5} \psi\rangle$, $a=2,5,7$. However, in the chiral 
limit (see the discussion in \cite{u2}) it is effectively depends on $M$, $\Pi$ and $\Delta$
quantities, where $M$ and $\Pi$ are the same as in the two-color case,
and $\Delta=\sqrt{|\Delta_2|^2+|\Delta_5|^2+|\Delta_7|^2}$. The above-established 
${\cal D_{\rm I}}$-invariance of each of the 3-color Lagrangians $L_{QCD}$ and 
$L_{NJL}$ means that the complete TDP in each of the models is symmetric with 
respect to the corresponding transformation of the order parameters and chemical potentials. 
It is easy to see that it looks like ${\cal D_{\rm I}}:M\leftrightarrow\Pi,~\mu_I\leftrightarrow\mu_{I5}$. 
Thanks to this circumstance, the study of the complete phase diagram of quark matter 
is greatly simplified, since CSB and charged PC phases should always be located dually 
conjugate, i.e. symmetrically, to each other in the phase portrait.
In addition, if we know, e.g., that at the point $(\mu_{B},\mu_{I}=A,\mu_{I5}=B,\mu_{5})$ a phase with color superconductivity (in which $\Delta\ne 0$, $M=\Pi=0$) is realized in the system, then, without any calculations, it can be argued that at the chemical potential point of the form $(\mu_{B},\mu_{I}=B,\mu_{I5}=A,\mu_{5})$ we observe the same phase.

{\bf Conclusions.---}Due to very 
extensive $SU(4)$ symmetry group of two-color QCD, 
it has been realized in our paper on the basis of formalism with auxiliary spinor fields $\Psi$ that there are
three different (duality) transformations (\ref{03})-(\ref{5})
of fields and chemical potentials such that $L_{QC_2D}$ and $L_{NJL_2}$ Lagrangians stay 
invariant. In this case fermion bilinear structures corresponding to the condensation 
phenomena transform between themselves. Hence, these duality transformations lead to the 
dual symmetries of the full phase diagram. One can mention that it is another example that 
consideration of two-color QCD is quite fruitful. 
Then we have seen that one of them, ${\cal D_{\rm I}}$, could be defined in three-color 
case and checked that this dual symmetry transformation takes place in this case as well, 
so studying two-color QCD we make an observation of three-color one.

After the proof of dual symmetries of QCD, they became the first principle method so rarely available in QCD phase diagram studies, and on top of that it is fully analytic method that does not need complicated calculation. It does not require any truncation or renormalization procedure and could not lead to any artifacts on its own or any other numeric complications. 

Dualities already proved to be rather useful tool for studying QCD phase diagram.
In three-color QCD duality connects isospin $\mu_I$ (without sign problem) and chiral 
$\mu_{I5}$ chemical potentials (present sign problem). So QCD at $\mu_{I5}$ is the only 
known example with sign problem, where full phase diagram could be obtained by dual mapping  
of the QCD at $\mu_I$, so maybe it might shed new light on solving sign problem and 
one can test different methods in real situation plagued by sign problem having the exact 
result. 

Duality in three color case and dualities ${\cal D_{\rm I}}$ and ${\cal D_{\rm III}}$ in two color one are exact only in the chiral limit, while duality ${\cal D_{\rm II}}$ in two color case is exact even at the physical point. It has been shown in \cite{khun2} that duality in three color NJL model in the mean field is a very good approximation at the physical point. Furthermore, all discussed dualities stay intact at any finite temperature.

Dual symmetries of QCD phase portrait have been shown from first principles and true 
at any energy scales, at any values of chemical potentials, 
as real-valued as well as imaginary ones, that could find some applications. 
Dualities can be easily used for obtaining new results, as it was already done in \cite{khun}. 
Moreover, because of them, we recently discovered that new inhomogeneous phases 
(at $\mu_B=0$ but with other nonzero chemical potentials) are
present both in three- and two-color QCD -- 
this result will be published elsewhere soon.

As dualities have been shown from first principles, all physical phenomena at one chemical potential are dually conjugated to the ones at other chemical potentials, including confinement/deconfinement phase transition (absent in NJL model), topological properties etc. Just to illustrate, let us discuss the following example.  There are indications that speed of sound squared could exceed the value $\frac{1}{3}$ (the conformal limit), the bound expected from holography some time ago \cite{Cherman:2009tw}, at nonzero baryon density \cite{Bedaque:2014sqa}. First example of this violation from first principle lattice QCD was obtained not at nonzero baryon but at nonzero isospin density, i.e. at $\mu_I\ne 0$, in 
\cite{brandt, abbott}. And by using duality one can immediately obtain without any (extremely time costly and even impossible in this case) numerical calculations that this bound is violated in quark matter with chiral imbalance at $\mu_{I5}\ne 0$ (also first principle result now). There was shown that this violation occurs also in two-color QCD at nonzero baryon chemical potential (see in \cite{itou}). By exploiting the rich duality structure of QC$_2$D, one can show that the speed of sound exceeds the conformal value at $\mu_I\ne 0$ and $\mu_{I5}\ne 0$ as well.
So maybe the violation by speed of sound of conformal bound in QCD is a natural phenomenon. 
In our opinion, these are only a few number of examples demonstrating that QCD dual symmetries are a fundamental tool for studying the properties of dense quark matter.


\begin{thebibliography}{}

\bibitem{buballa2}
S.~P.~Klevansky,
Rev.\ Mod.\ Phys.\  {\bf 64}, 649 (1992).

\bibitem{asakawa}
M.~Asakawa and K.~Yazaki,
Nucl. Phys. A \textbf{504}, 668 (1989);
P.~Zhuang, J.~Hufner and S.~P.~Klevansky,
Nucl. Phys. A \textbf{576}, 525 (1994).

\bibitem{alford}
M. Buballa, Phys. Rep. {\bf 407}, 205 (2005);
I. A. Shovkovy, Found. Phys. {\bf 35}, 1309 (2005);
M. G.~Alford, A.~Schmitt, K.~Rajagopal, and T.~Sch\"afer,
 Rev.\ Mod.\ Phys.\  {\bf 80}, 1455 (2008);
E.~J.~Ferrer and V.~de la Incera,
Lect.\ Notes Phys.\  {\bf 871}, 399 (2013).

\bibitem{Vilenkin}
A.~Vilenkin,
Phys. Rev. D \textbf{22}, 3080-3084 (1980); M.~A.~Metlitski and A.~R.~Zhitnitsky,
Phys. Rev. D \textbf{72}, 045011 (2005).

\bibitem{Shovkovy}
I.~A.~Shovkovy,
``Anomalous plasma: chiral magnetic effect and all that,''
[arXiv:2111.11416 [nucl-th]].

\bibitem{Ruggieri:2016lrn}
M.~Ruggieri and G.~X.~Peng,
Phys. Rev. D \textbf{93} (2016) no.9, 094021;
M.~Ruggieri, G.~X.~Peng and M.~Chernodub,
EPJ Web Conf. \textbf{129} (2016), 00037

\bibitem{kkz20}
T.~G.~Khunjua, K.~G.~Klimenko and R.~N.~Zhokhov,
JHEP \textbf{06}, 148 (2020);
Phys. Part. Nucl. \textbf{53}, no.2, 461 (2022);
Phys. Rev. D \textbf{106}, no.4, 045008 (2022).

\bibitem{khun2}
T.~G.~Khunjua, K.~G.~Klimenko and R.~N.~Zhokhov,
Eur. Phys. J. C \textbf{79}, no.2, 151 (2019); 
J. Phys. Conf. Ser. \textbf{1390}, no.1, 012015 (2019).

\bibitem{khun}
T.~G.~Khunjua, K.~G.~Klimenko and R.~N.~Zhokhov,
JHEP \textbf{06}, 006 (2019).

\bibitem{Fukushima}
K.~Fukushima, D.~E.~Kharzeev and H.~J.~Warringa,
Phys. Rev. D \textbf{78}, 074033 (2008).

\bibitem{u2}
T.~G.~Khunjua, K.~G.~Klimenko and R.~N.~Zhokhov,
Phys. Rev. D \textbf{108}, no.12, 125011 (2023).

\bibitem{son}
D. T.~Son and M. A.~Stephanov, Phys.\ Atom.\ Nucl.\  {\bf 64}, 834 (2001);
D. C.~Duarte, R. L. S.~Farias and R. O.~Ramos,
Phys.\ Rev.\  D {\bf 84}, 083525 (2011).

\bibitem{he}
L. He, M. Jin, and P. Zhuang, Phys. Rev. D {\bf 71}, 116001 (2005);
D. Ebert and K. G. Klimenko, J.\ Phys.\ G {\bf 32}, 599 (2006);
Eur.\ Phys.\ J.\  C {\bf 46}, 771 (2006);
C.f.~Mu, L.y.~He and Y.x.~Liu,
  Phys.\ Rev.\  D {\bf 82}, 056006 (2010).

\bibitem{ak}
J. O.~Andersen and T.~Brauner,
  Phys.\ Rev.\  D {\bf 78}, 014030 (2008);
J. O.~Andersen and L.~Kyllingstad,
 J.\ Phys.\ G {\bf 37}, 015003 (2009);
P.~Adhikari, J.~O.~Andersen and P.~Kneschke,
Eur.\ Phys.\ J.\ C {\bf 79},  874 (2019).

\bibitem{ekkz2}
 D.~Ebert, T. G.~Khunjua, K. G.~Klimenko and V. C.~Zhukovsky,
  Int.\ J.\ Mod.\ Phys.\ A {\bf 27}, 1250162 (2012);
  N. V.~Gubina, K. G.~Klimenko, S. G.~Kurbanov and V. C.~Zhukovsky,
  Phys.\ Rev.\ D {\bf 86}, 085011 (2012).

  \bibitem{andrianov}
 R.~Gatto and M.~Ruggieri,
Phys.\ Rev.\ D {\bf 85}, 054013 (2012);
 L.~Yu, H.~Liu and M.~Huang,
Phys.\ Rev.\ D {\bf 90}, 074009 (2014);
 M.~Ruggieri and G.~X.~Peng,
J.\ Phys.\ G {\bf 43}, no. 12, 125101 (2016);
A.~A.~Andrianov, V.~A.~Andrianov and D.~Espriu,
Particles {\bf 3}, no. 1, 15 (2020).

\bibitem{chao}
  J.~Chao, Chin. Phys. C \textbf{44}, no.3, 034108 (2020).
  
\bibitem{ramos}
F.~X.~Azeredo, D.~C.~Duarte, R.~L.~S.~Farias, G.~Krein and R.~O.~Ramos,
arXiv:2406.04900 [hep-ph].

\bibitem{kkz18}
T.~G.~Khunjua, K.~G.~Klimenko and R.~N.~Zhokhov,
Phys. Rev. D \textbf{97}, no.5, 054036 (2018).

\bibitem{kkz18-2}
T.~G.~Khunjua, K.~G.~Klimenko and R.~N.~Zhokhov,
Phys. Rev. D \textbf{98}, no.5, 054030 (2018).

\bibitem{kkzparticles} 
T.~G.~Khunjua, K.~G.~Klimenko and R.~N.~Zhokhov,
Particles {\bf 3}, no. 1, 62 (2020).

\bibitem{k}
T.~G.~Khunjua, K.~G.~Klimenko and R.~N.~Zhokhov,
Phys. Rev. D \textbf{94}, no.11, 116016 (2016).

\bibitem{kkz}
T.~G.~Khunjua, K.~G.~Klimenko and R.~N.~Zhokhov,
Phys. Rev. D \textbf{100}, no.3, 034009 (2019);
M.~Thies, 
Phys. Rev. D \textbf{101}, no.1, 014010 (2020);
Phys. Rev. D \textbf{102}, no.9, 096006 (2020).

\bibitem{thies}
M.~Thies,
Phys. Rev. D \textbf{68}, 047703 (2003);
Phys. Rev. D \textbf{90}, no.10, 105017 (2014).

\bibitem{ebert}
D.~Ebert, T.~G.~Khunjua, K.~G.~Klimenko and V.~C.~Zhukovsky,
Phys. Rev. D \textbf{90}, no.4, 045021 (2014);
Phys. Rev. D \textbf{93}, no.10, 105022 (2016).

\bibitem{cao}
G.~Cao, L.~He and P.~Zhuang,
Phys. Rev. D \textbf{90}, no.5, 056005 (2014).

\bibitem{Kogut}
J.~B.~Kogut, M.~A.~Stephanov and D.~Toublan,
Phys. Lett. B \textbf{464}, 183-191 (1999);
J.~B.~Kogut, M.~A.~Stephanov, D.~Toublan, J.~J.~M.~Verbaarschot and A.~Zhitnitsky,
Nucl. Phys. B \textbf{582}, 477 (2000).

\bibitem{SonSplittorff}
K.~Splittorff, D.~T.~Son and M.~A.~Stephanov,
Phys. Rev. D \textbf{64}, 016003 (2001).

\bibitem{Andersen}
J.~O.~Andersen and T.~Brauner,
Phys. Rev. D \textbf{81}, 096004 (2010).

\bibitem{Cherman:2009tw}
A.~Cherman, T.~D.~Cohen and A.~Nellore,
Phys. Rev. D \textbf{80} (2009), 066003

\bibitem{Bedaque:2014sqa}
P.~Bedaque and A.~W.~Steiner,
Phys. Rev. Lett. \textbf{114} (2015) no.3, 031103;
I.~Tews, J.~Carlson, S.~Gandolfi and S.~Reddy,
Astrophys. J. \textbf{860} (2018) no.2, 149;
S.~Altiparmak, C.~Ecker and L.~Rezzolla,
Astrophys. J. Lett. \textbf{939} (2022) no.2, L34

\bibitem{brandt}
B.~B.~Brandt, F.~Cuteri and G.~Endrodi,
JHEP \textbf{07}, 055 (2023).

\bibitem{abbott}
R.~Abbott \textit{et al.} [NPLQCD],
Phys. Rev. D \textbf{108}, no.11, 114506 (2023).

\bibitem{itou}
E.~Itou and K.~Iida,
PoS \textbf{LATTICE2023}, 111 (2024); 
PTEP \textbf{2022} (2022) no.11, 111B01

\end{thebibliography}
\end{document}